\documentclass{article}[11pt]	
\title{Approximating the Turaev-Viro Invariant of Mapping Tori is
  Complete for One Clean Qubit}
\author{Stephen P. Jordan\footnote{Current
    affiliation: NIST (\emph{stephen.jordan@nist.gov}). This work was done at
    Institute for Quantum Information, Caltech.} \ and Gorjan
  Alagic\footnote{Institute for Quantum Computation, University of
    Waterloo (\emph{galagic@iqc.ca}).}}
\date{}

\usepackage{graphicx}
\usepackage{amssymb}
\usepackage{amsmath}
\usepackage{fullpage}
\usepackage{dsfont}

\input{Qcircuit}

\newcommand{\captionfonts}{\small}

\makeatletter  
\long\def\@makecaption#1#2{%
  \vskip\abovecaptionskip
  \sbox\@tempboxa{{\captionfonts #1: #2}}%
  \ifdim \wd\@tempboxa >\hsize
    {\captionfonts #1: #2\par}
  \else
    \hbox to\hsize{\hfil\box\@tempboxa\hfil}%
  \fi
  \vskip\belowcaptionskip}
\makeatother   

\begin{document}
\bibliographystyle{plain}
\maketitle
\newcommand{\ud}{\mathrm{d}}
\newcommand{\Bra}[1]{\left<#1\right|}
\newcommand{\Ket}[1]{\left|#1\right>}
\newcommand{\Braket}[2]{\left< #1 \right| #2 \right>}
\renewcommand{\th}{^\mathrm{th}}
\newcommand{\tr}{\mathrm{Tr}}
\newcommand{\id}{\mathds{1}}

\def\identity{\mathds 1}
\def\CC{\mathbb C}
\def\ZZ{\mathbb Z}
\def\RR{\mathbb R}
\def\PP{\mathbb P}

\newtheorem{lemma}{Lemma}
\newtheorem{theorem}{Theorem}
\newtheorem{prop}{Proposition}

\begin{abstract}
The Turaev-Viro invariants are scalar topological invariants of
three-dimensional manifolds. Here we show that the problem of
estimating the Fibonacci version of the Turaev-Viro invariant of a
mapping torus is a complete problem for the one clean qubit complexity
class (DQC1). This complements a previous result showing that
estimating the Turaev-Viro invariant for arbitrary manifolds presented
as Heegaard splittings is a complete problem for the standard quantum
computation model (BQP). We also discuss a beautiful analogy between
these results and previously known results on the computational
complexity of approximating the Jones Polynomial.
\end{abstract}

\section{Introduction}

Classifying the power of quantum computers is a fundamental problem in
quantum information science. The computational power of a
general-purpose quantum computer is identified with the complexity
class BQP (bounded-error quantum polynomial time). The famous problems
of factoring and discrete logarithm, for instance, are in BQP. An
essential ingredient of BQP computation is the ability to initialize a
large number of qubits into a specific pure state. In some proposed
physical implementations, however, this appears to be an extremely
difficult task. In 1998, Knill and Laflamme proposed that exponential
speedups over classical computers could still be possible, even if one
can only initialize a single qubit into a pure state, with the rest of
the qubits in the maximally mixed state~\cite{Knill_Laflamme98}. The
complexity class thus defined is called DQC1 (deterministic quantum
computation with one clean qubit), or simply ``the one clean qubit
class.'' This class contains several problems for which no efficient
classical algorithms are known. The most basic of these is the problem
of estimating the trace of a unitary operator. In fact, trace
estimation is DQC1-complete: not only is it in DQC1, but any other
problem in DQC1 can be reduced to it.

Finding natural BQP-complete and DQC1-complete problems is essential
to our understanding of the computational power afforded by quantum
computers. Remarkably, BQP-complete problems can be found in areas of
mathematics without \emph{a priori} close connection to quantum
computation. In particular, approximating the Jones polynomial, a
famous invariant of links, is a BQP-complete problem~\cite{Freedman2,
  Freedman1, FLW02, Aharonov_Arad, AJL, Wocjan_Yard}. The input is an
element of the braid group, and the output is an estimate of the Jones
polynomial of the so-called \emph{plat closure} of the
braid. Estimating the Jones polynomial of the so-called \emph{trace
  closure} of the braid is DQC1-complete~\cite{Shor_Jordan,
  Jordan_Wocjan}.

Recent work~\cite{AJKR, Garnerone} showed that (the decision version of)
approximating certain invariants of 3-manifolds is a BQP-complete
problem. In this formulation, the input is a so-called \emph{Heegaard
  splitting} of a 3-manifold, specified as an element of the mapping
class group. The output is an estimate of the Turaev-Viro invariant of
the input manifold. In this article we show that approximating the
Turaev-Viro invariant of a 3-manifold specified as a \emph{mapping
  torus} is a complete problem for the one clean qubit class. In
section \ref{sec:analogy}, we use the language of Topological
Quantum Field Theories (or TQFTs) to explain the mathematical
underpinnings of the relationship between approximating the Jones
polynomial of the plat and trace closures, and approximating the
Turaev-Viro invariant of Heegaard splittings and mapping tori.

We assume only a basic understanding of topology and quantum
computation. Needed concepts in manifold invariants and one clean
qubit computation are explained in section \ref{background}.
Our exposition focuses on the Witten-Reshetikhin-Turaev (or WRT)
invariant. This is only a matter of convenience, as it is known that 
the Turaev-Viro invariant is equal to the absolute square of the WRT
invariant~\cite{Turaev91, TuraevBook, Walker, Roberts}. 

\section{Background}
\label{background}

\subsection{Two-manifolds and three-manifolds}

We begin by setting down a few basic definitions from low-dimensional
topology. Recall that an \emph{$n$-manifold} is a topological
space\footnote{more precisely, a second-countable Hausdorff space}
whose every point has a neighborhood that looks like (i.e., is
homeomorphic to) an open subset of $\RR^n$. Simple examples of
one-dimensional manifolds include the line $\RR$ and the circle $S^1$.
Simple examples of two-dimensional manifolds include the
the plane $\RR^2$, the sphere $S^2$, and the torus $\Sigma_1 = S^1 \times
S^1$, which we can visualize as the surface of a donut. More generally,
the surface of a donut with $g$ holes is also a two-manifold, which we call
the surface of genus $g$ and denote by $\Sigma_g$. The genus is a
complete invariant of surfaces\footnote{In this work, we implicitly
  assume that all surfaces are closed, compact, connected and
  orientable.}: homeomorphic surfaces have the same number of handles
(invariance), and non-homeomorphic surfaces have a different number of
handles (completeness).

The simplest example of a $3$-manifold is $\RR^3$ itself. A nontrivial
example is found by taking the product of $\Sigma_1$ with a third
circle; the result is the three-dimensional torus $T^3 = S^1 \times S^1
\times S^1$. Given a surface $\Sigma_g$, the \emph{cylinder} $\Sigma_g
\times [0,1]$ is a $3$-manifold whose boundary consists of two copies
of $\Sigma_g$ (specifically, the bottom $\Sigma_g \times \{0\}$ and
the top $\Sigma_g \times \{1\}$.) We can turn the cylinder into a
$3$-manifold without boundary by choosing a homeomorphism $f: \Sigma_g
\rightarrow \Sigma_g$ and gluing each point on the top to its image
under $f$ on the bottom. The result is the \emph{mapping torus} of
$f$:
$$
T_{g, f} = \frac{\Sigma_g \times [0,1]}{(x,1) \sim (f(x),0)}~.
$$
For example, choosing $g = 1$ and $f$ to be the identity map, we see
that $T_{1, \identity} = T^3$. A useful example of a nontrivial
self-homeomorphism of $\Sigma_g$ is the so-called Dehn twist. To visualize a
Dehn twist, imagine cutting the handle of $\Sigma_1$ to get a tube,
performing a $2\pi$ twist on one end of the tube, and then gluing the
handle back together. In general, a Dehn twist can be performed around
any noncontractible closed curve.

\begin{figure}
\begin{center}
\includegraphics[width=0.5\textwidth]{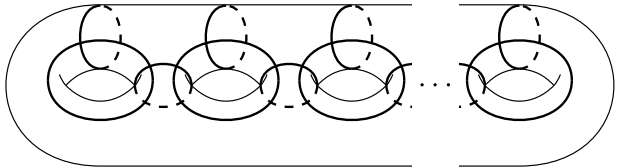}
\caption{\label{cancurves} A Dehn twist is a $2 \pi$ rotation about a
  closed curve. The Dehn twists along the $3g-1$ curves illustrated
  here constitute a standard set of generators for the mapping class
  group of the genus $g$ surface.}
\end{center}
\end{figure}

The (homeomorphism class of) the mapping torus $T_{g,f}$ depends only
on the isotopy class of $f$. The orientation-preserving
self-homeomorphisms of $\Sigma_g$ form a group under composition. This
group, taken modulo isotopy, is called the mapping class group of
$\Sigma_g$, and is denoted $\mathrm{MCG}(g)$. $\mathrm{MCG}(g)$ is
generated by the Dehn twists about the $3g-1$ canonical curves shown
in figure \ref{cancurves}. Any mapping torus $T_{g, f}$ is thus
described by a word in the Dehn twist generators of
$\mathrm{MCG}(g)$.

\subsection{The Witten-Reshetikhin-Turaev invariants}\label{WRTdefinition}

Recall that the genus is an invariant of surfaces because it assigns
the same number to homeomorphic surfaces. One can also define
invariants of $3$-manifolds, although none are as simple and powerful
as the genus. In the 1990s, Witten, Reshetikhin, and Turaev discovered
a family of $3$-manifold invariants arising from their work in
Topological Quantum Field Theory. While these invariants can be
defined for arbitrary $3$-manifolds, we only concern ourselves
with the special case of mapping tori, where the definitions are
relatively straightforward. Specifically, the
Witten-Reshetikhin-Turaev (WRT) invariant of a mapping torus $T_{g,
  f}$ is equal to the trace of $f$ in a certain projective
representation of the mapping class group $\mathrm{MCG}(g)$. Note that
the WRT function is only a topological invariant up to a phase
(see~\cite{AJKR}). In general, the WRT invariant is parametrized by a
quantum group, such as $\mathrm{SU}(N)_k$ or
$\mathrm{SO}(N)_k$. Although some of our results apply more generally,
we focus on the case of $\mathrm{SO}(3)_3$, sometimes called the
Fibonacci model. In this case, the description of the representation
is particularly simple, and can be understood with no background in
quantum groups.

\begin{figure}
\begin{center}
\includegraphics[width=0.6\textwidth]{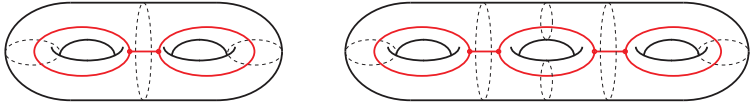}
\caption{\label{spine} The dashed lines indicate a set of cuts that
  decomposes the surface into two three-punctured-spheres
  (``pants''). Dual to this is a trivalent graph called the ``spine,''
  in red. The genus two and genus three cases are shown here.}
\end{center}
\end{figure}

\begin{figure}
\begin{center}
\includegraphics[width=\textwidth]{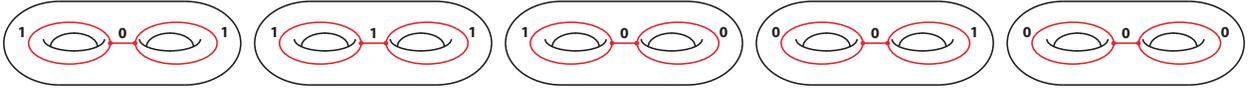}
\caption{\label{genustwo} The Fibonacci model's fusion rules allow
  five labelings of the standard spine of the genus two surface. This
  means that the WRT representation of $\mathrm{MCG}(2)$ is
  five-dimensional.}
\end{center}
\end{figure}

The Fibonacci representation is defined as follows. Any genus $g$
surface (for $g > 1)$ can be cut into three-punctured spheres,
resulting in a so-called pants decomposition. Dual to such a
decomposition is a trivalent graph on the surface, called a spine. As
illustrated in figure \ref{spine}, the spine has one vertex for every
pant in the decomposition. Whenever two pants meet at a puncture, the
spine has an edge between the corresponding vertices. While a surface
admits many spines (and corresponding pants decompositions), we call
the one shown in figure \ref{spine} the \emph{standard spine}. We
label the edges of the standard spine by so-called anyon types,
with fusion rules enforced at each vertex. For the Fibonacci model,
there are only two anyon types: $0$ and $1$, and only one fusion rule:
no vertex can have exactly two edges labeled $0$ incident on it. The
case $g=2$ is pictured in figure \ref{genustwo}. The formal span (over
$\CC$) of all such labelings associates a finite-dimensional vector
space to the surface. Different spines yield different bases for this
same space. We can move between these spines (and the corresponding
bases) by means of two ``moves,'' the F-move:
\[
\begin{array}{c}
\includegraphics[width=0.13\textwidth]{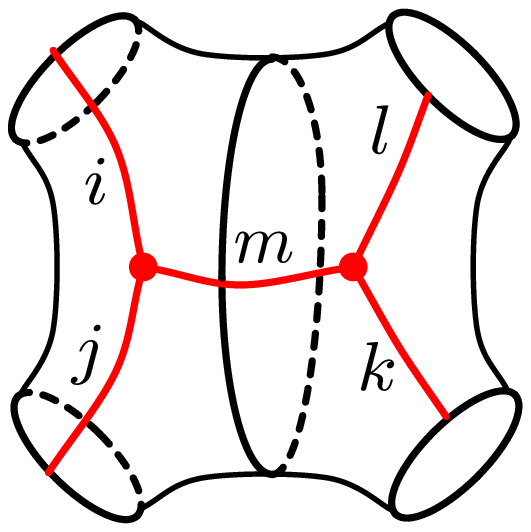}
\end{array}
= \sum_n F^{i j m}_{k l n} 
\begin{array}{c}
\includegraphics[width=0.13\textwidth]{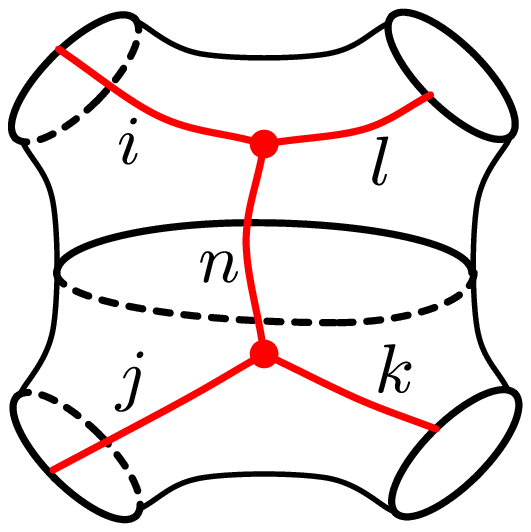}
\end{array}
\]
and the S-move:
\[
\begin{array}{c}
\includegraphics[width=0.13\textwidth]{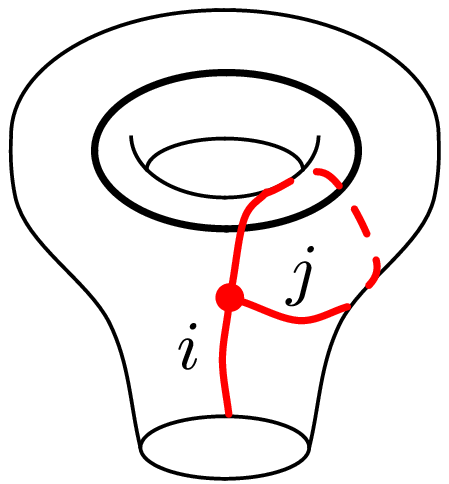}
\end{array}
= \sum_k S^i_{jk} 
\begin{array}{c}
\includegraphics[width=0.13\textwidth]{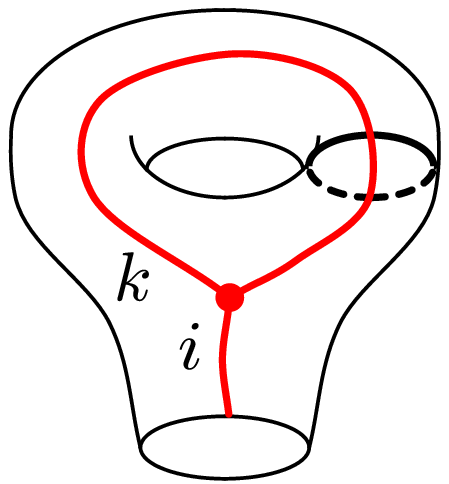}
\end{array}
\]
For the Fibonacci model $F_{abc}^{def}$ is as follows
\begin{center}
\includegraphics[width=0.5\textwidth]{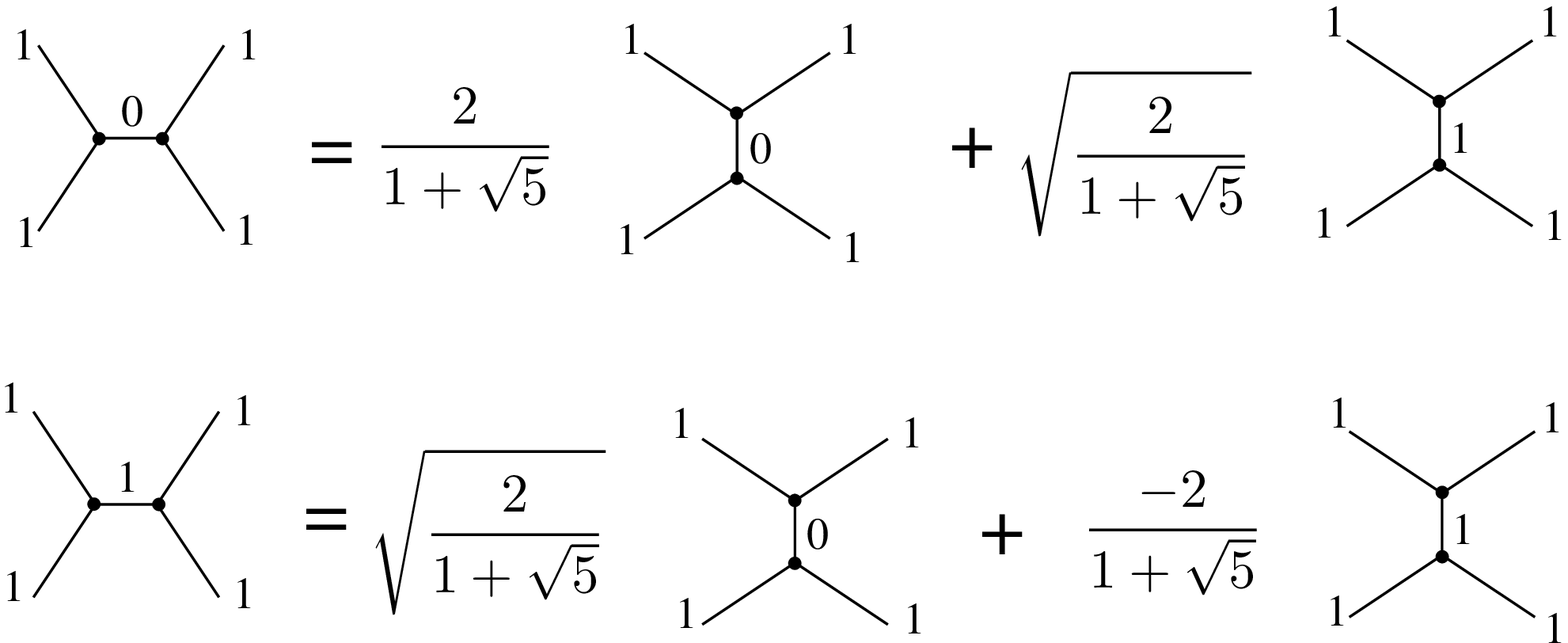}
\end{center}
with all other values equal to zero or one as dictated by the fusion
rules. As one can calculate using the prescription described
in~\cite{AJKR}, $S^i_{jk}$ is given in the Fibonacci model by
\begin{eqnarray*}
D S^0_{00} & = & 1 \\
D S^0_{10} = D S^0_{01} & = & \frac{1+\sqrt{5}}{2} \\
D S^0_{11} & = & 1 + \frac{1+ \sqrt{5}}{2} e^{i 4 \pi/5} \\
D S^1_{11} & = & \sqrt{\frac{1+\sqrt{5}}{2}} \left(1 - e^{i 4 \pi/5}
\right)
\end{eqnarray*}
with $D = \sqrt{1+\left(\frac{1+\sqrt{5}}{2} \right)^2}$ and all other
values of $S^i_{jk}$ equal to zero by the fusion rules.

The space described above is the underlying vector space for the
Fibonacci representation of $\mathrm{MCG}(g)$. We define this
representation in the basis corresponding to the standard spine. Since
the mapping class group is finitely-generated, it suffices to describe
the images of the Dehn twist generators. Any such generator is a $2
\pi$ twist along some canonical curve $c$ from figure
\ref{cancurves}. It is not hard to check that, by applying at most one
F-move and one S-move, the standard spine can be adjusted so that $c$
is a cut in the corresponding pants decomposition. In this basis, the
Dehn twist about $c$ induces a diagonal linear transformation. To each
labeling of the spine corresponds a basis vector, and this basis
vector obtains a phase determined by the label on the edge of the
spine that intersects $c$. In the Fibonacci model, edges labeled 0
obtain a phase of 1, and edges labeled 1 obtain a phase of $e^{i 3
  \pi/5}$. In the standard spine basis, the matrix corresponding to
the Dehn twist about $c$ is thus simply a product of at most five
matrices: at most two of the moves pictured above, followed by a
diagonal matrix, followed by the inverse moves to return to the
original basis. The WRT invariant of the mapping torus $T_{g, f}$ is
now simply the trace of the Fibonacci representation, evaluated at
$f$.

\subsection{One Clean Qubit}

In some proposed implementations of quantum computers, such as nuclear
magnetic resonance (NMR) the most difficult task is initializing
qubits into a pure state. In 1998, Knill and Laflamme proposed that
exponential speedups over classical computation might be possible
without pure state initialization. To mathematically investigate this
possibility, they introduced the one clean qubit
model~\cite{Knill_Laflamme98}. In this model, one is given an initial
state $\rho$ with $n$ qubits in the maximally mixed state, and one
qubit in the pure state $\ket{0}$.
\[
\rho = \ket{0} \bra{0} \otimes \frac{\id}{2^n}
\]
One then applies any quantum circuit of $\mathrm{poly}(n)$ gates to
this state, and measures the first qubit in the computational
basis. Computational problems are solved by performing polynomially
many such experiments, each starting with the initial state $\rho$,
and recording the output statistics. The class of decision problems
solvable with bounded probability of error using this procedure is
called DQC1.

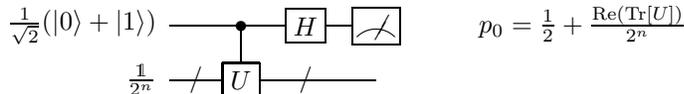
\begin{figure}
\begin{center}
\[
\begin{array}{lll}
\Qcircuit @C=1em @R=.7em {
  \lstick{\frac{1}{\sqrt{2}}(\ket{0}+\ket{1})} & \qw     & \ctrl{1} & \gate{H} & \meter \\
  \lstick{\frac{\id}{2^n}}                     & {/} \qw & \gate{U} & {/} \qw  & \qw} 
  & \quad & p_0 = \frac{1}{2}+\frac{\mathrm{Re}(\tr [U])}{2^n}
\end{array}
\]
\caption{\label{trace} By repeating this one clean qubit computation,
  and recording the fraction of $0$ outcomes, one estimates the real
  part of $\tr [U]/2^n$. Similarly, by initializing the clean qubit to 
  $\frac{1}{\sqrt{2}} (\ket{0} - i \ket{1})$, one obtains 
  $p_0 = \frac{1}{2}+\frac{\mathrm{Im}(\tr [U])}{2^n}$.}
\end{center}
\end{figure}

DQC1 contains several computational problems not known to be solvable
in polynomial time on classical computers. Most fundamentally, given a
description of a quantum circuit of $T$ gates implementing the unitary
transformation $U$ on $n$ qubits, a one clean qubit computer can
estimate the normalized trace $\frac{\tr{U}}{2^n}$ to within $\pm
\epsilon$ in time $O(T/\epsilon^2)$ by means of the circuit shown in
figure \ref{trace}. Furthermore, this problem of estimating the trace
of a quantum circuit is DQC1-hard~\cite{Knill_Laflamme98, Shepherd,
  Shor_Jordan}.  Efficient one clean qubit algorithms have been
discovered for estimating certain quadratically signed weight
enumerators~\cite{Knill_Laflamme01} and estimating certain
Jones~\cite{Shor_Jordan} and HOMFLY~\cite{Jordan_Wocjan}
polynomials. A version of the Jones polynomial problem is
DQC1-complete~\cite{Shor_Jordan}, and has been demonstrated
experimentally with NMR~\cite{Passante, Marx}. A certain problem of
approximating partition functions for quantum systems is also
DQC1-hard~\cite{Brandao}.

In many ways, it is surprising that one clean qubit computers can do
any nontrivial computations at all. If all $n+1$ qubits were maximally
mixed, the resulting state would be invariant under all unitaries.
Furthermore, DQC1 computations involve very
little entanglement~\cite{Datta_thesis, Datta_Gharibian,
  Datta_Flammia_Caves, Datta_Shaji_Caves, Datta_Vidal, Luo}. Ambainis
\emph{et al.} give an impossibility proof against a certain natural
approach to simulating standard quantum computers using one clean
qubit computers, and on the other hand show that one clean qubit
computers can efficiently simulate classical logarithmic
depth (NC1) computations~\cite{Ambainis_Schulman_Vazirani}.

The DQC1 complexity class is robust against a variety of modifications
to the computational model. The class of computational problems
solvable in polynomial time with up to logarithmically many clean
qubits is the same as that solvable in polynomial time with one clean
qubit~\cite{Shor_Jordan}. If the clean qubit is not pure,
but has $1/\mathrm{poly}(n)$ polarization, the set of efficiently
solvable problems also remains DQC1~\cite{Knill_Laflamme98}. As shown
in appendix \ref{qudits}, the one clean qudit model on $d$-dimensional
qudits is equivalent in power to the one clean qubit model, for any
constant $d$.

\section{Algorithm}
\label{algorithm}

In this section we construct an efficient one clean qubit algorithm
for approximating the Fibonacci WRT invariant of a mapping
torus. Generalizing to other tensor categories such as
$\mathrm{SU}(N)_k$ and $\mathrm{SO}(N)_k$ is straightforward. The main
idea of the algorithm is, given a word $w$ in the Dehn twist
generators of $\mathrm{MCG}(g)$, to find a quantum circuit of
$\mathrm{poly}(w,g)$ gates on $\mathrm{poly}(g)$ qubits whose trace is
equal to the WRT invariant of the $3$-manifold $T_{g, w}$. This trace
can then be approximated by means of the circuit in figure
\ref{trace}. For this purpose, we encode the allowed labelings of a
spine of $\Sigma_g$ into qubits, and then construct a quantum circuit
implementing the Fibonacci representation of $\mathrm{MCG}(g)$ on this
encoding. The most obvious encoding would be to directly assign one
qubit to store the particle type for each edge of the spine. However,
a one clean qubit computer yields the normalized trace over all $2^n$
bitstrings, of which only an exponentially small fraction represent
valid spine labelings in this encoding.

We instead construct a many-to-one map $\varphi:\{0,1\}^{\beta (3g-3)} \to
\{ \textrm{valid labelings} \}$ with $\beta = O(\log |g|)$ such that the
preimage of each spine-labeling consists of approximately the same
number of bitstrings. That is, $|\varphi^{-1}(x)|$ is approximately
independent of $x$. Thus, the normalized trace of the Fibonacci
representation of $w \in \mathrm{MCG}(g)$ acting on the
$\varphi$-encoded labelings of the spine of $\Sigma_g$ is
approximately equal to $\mathrm{WRT}(T_{g, w})$. We construct
$\varphi$ following a method introduced in~\cite{Jordan_Wocjan}. We
assign a register of $\beta =  O(\log |g|)$ qubits to each edge of the
spine. The bitstring contained in register $i$ is interpreted as an
integer $0 \leq x_i \leq 2^\beta  -1$. We then assign a threshold
$T_i$ so that $x_i \leq T_i$ encodes a zero label on edge $i$, and
$x_i > T_i$ encodes a one label. By carefully choosing the thresholds
$T_1,\ldots,T_{3g-3}$ we ensure that $| \varphi^{-1}(x) |$ is
approximately independent of $x$.

\begin{figure}
\begin{center}
\includegraphics[width=0.3\textwidth]{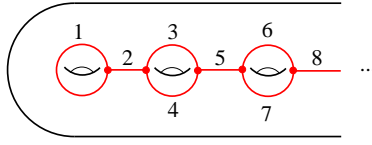}
\caption{\label{numbering}We number the edges of the standard spine
  from left to right, with ambiguities resolved by ordering from top
  to bottom.}
\end{center}
\end{figure}

Number the edges of the spine from one to $3g-3$, left to right and
top to bottom, as illustrated in figure \ref{numbering}. Let
$s_1,\ldots,s_{3g-3} \in \{0,1\}^{3g-3}$ be the labels of these
edges. The uniform probability distribution over all fusion-consistent
labelings of the spine induces a probability distribution
$p_g(s_1\ldots,s_{3g-3})$ over $\{0,1\}^{3g-3}$, with zero
probability for strings that violate fusion rules, and uniform
probability for the rest. For the genus-$g$ standard spine, we define
$p_g(s_i|s_1,\ldots,s_{i-1})$ to be the conditional probability that
label $i$ takes the value $s_i$ given that labels 1 through $i-1$ take
the values $s_1, \ldots, s_{i-1}$. For a register representing a label
$s_i$ we choose the threshold dependent on the values of
$s_1,\ldots,s_{i-1}$ according to 
\begin{equation}
\label{threshold}
T_i(g;s_1,\ldots,s_{i-1}) = \left \lceil 2^\beta p_g(0|s_1,\ldots,s_{i-1}) \right \rfloor.
\end{equation}
One can see that this choice ensures that a 
uniformly selected assignment of bitstrings to the registers yields a
uniform distribution over fusion-consistent labelings, up to the
errors induced by rounding. Hence, $|\varphi^{-1}(x)|$ is
approximately independent of $x$. More precisely, let
\begin{eqnarray*}
\tilde{p}_g(0|s_1,\ldots,s_{i-1}) & = & T_i(g;s_1,\ldots,s_{i-1})/2^\beta\\
\tilde{p}_g(1|s_1,\ldots,s_{i-1}) & = & 1 - \tilde{p}_g(0|s_1,\ldots,s_{i-1})
\end{eqnarray*}
Thus,
\begin{eqnarray*}
|\varphi^{-1}(s_1,\ldots,s_{3g-3})| & = & 2^{\beta (3g-3)}
  \tilde{p}_g(s_{3g-3}|s_1,\ldots,s_{3g-4}) \times
  \tilde{p}_g(s_{3g-4}|s_1,\ldots,s_{3g-5}) \times \ldots \times p(s_1)
  \\
& = & 2^{\beta (3g-3)}
  \left( p_g(s_{3g-3}|s_1,\ldots,s_{3g-4}) \pm O(2^{-\beta}) \right) 
\times \ldots \times \left( p_g(s_1) \pm O(2^{-\beta}) \right) \\
& = & p_g(s_1,\ldots,s_{3g-3}) \pm O(g 2^{-\beta}).
\end{eqnarray*}
Thus it suffices to choose $\beta = O( \log g)$. Furthermore, by the
locality of the fusion rules, $p_g(s_i|s_1,\ldots,s_{i-1})$ is always
independent of $s_1,\ldots,s_{i-3}$.  We may thus write
\begin{eqnarray}
p_g(s_i|s_1,\ldots,s_{i-1}) & = & p_g(s_i,s_{i-1},s_{i-2};i) \nonumber\\
T_i(g;s_1,\ldots,s_{i-1}) & = & T_i(g;s_i,s_{i-1},s_{i-2}). \label{independence}
\end{eqnarray}

\begin{figure}
\begin{center}
\[
\begin{array}{rl}
\begin{array}{c} \includegraphics[width=0.17\textwidth]{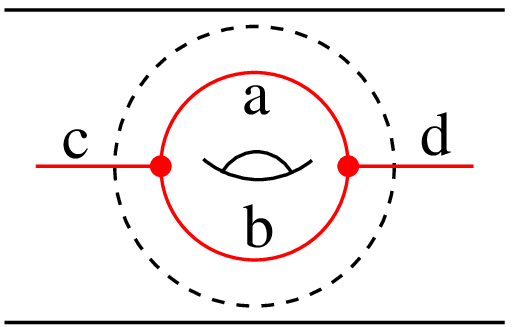}
\end{array} & \textrm{acts on $a,b,c,d$} \\
\begin{array}{c}
\includegraphics[width=0.17\textwidth]{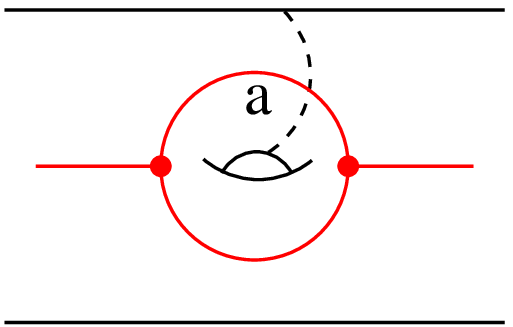} 
\end{array} & \textrm{acts on $a$} \\
\begin{array}{c} \includegraphics[width=0.25\textwidth]{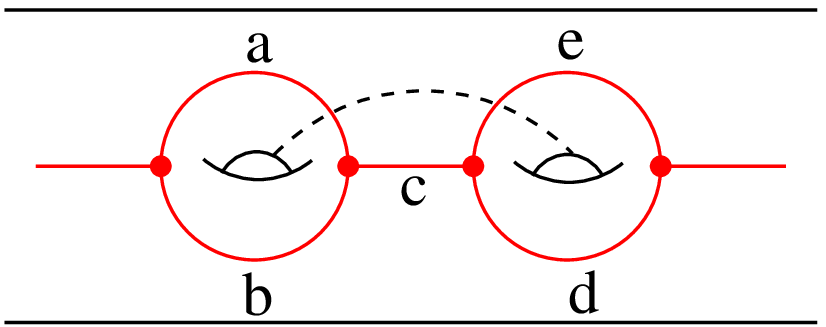} 
\end{array} & \textrm{acts on $a,b,c,d,e$ }
\end{array}
\]
\caption{\label{fivelabels}The Fibonacci representation of a Dehn
  twist (shown as a dashed line) from the standard generating set is a
  unitary transformation acting on at most five spine labels.}
\end{center}
\end{figure}

As illustrated in figure \ref{fivelabels}, the Fibonacci
representation of a Dehn twist from the standard generating set is a
unitary transformation acting on at most five spine labels. Because
the encoding $\varphi$ is many-to-one, the unitary transformation on
these spine labels does not uniquely define a unitary operation on the
bitstrings encoding them. We say that a pair of spine-labelings is
\emph{connected} if the Fibonacci representation of a Dehn twist from
the standard set of generators has a nonzero matrix element between
them. By choosing a bijection $b_{x,y}$ between the encodings of each pair
of connected spin-labelings we define a unitary transformation on the
encodings: if the matrix element between labeling $x$ and $y$ is
$\rho_{x,y}$ then, 
\begin{equation}
\label{udef}
U_{i,j}  = \left\{ \begin{array}{ll} \rho_{x,y} & \textrm{if $\varphi(i) =
    x$, $\varphi(j) = y$, and $b_{x,y}(i) = j$} \\
0 & \textrm{otherwise}
\end{array} \right.
\end{equation}
is a corresponding unitary representation on the encodings. Our choice
of bijections does not matter. We may for concreteness match
bitstrings by lexicographic ordering. One can verify that $U$ is a
direct sum of many copies of the Fibonacci representation
$\rho$. (The rounding involved in (\ref{threshold}) introduces a minor
technical complication, whose resolution may be found in
\cite{Jordan_Wocjan}.)

For any of the standard Dehn twist generators, $U_{i,j}$ acts on at
most $5 \beta$ qubits, which encode the spine-labels on which $\rho$
acts. The matrix elements by which $U$ acts on these qubits depends on
the corresponding thresholds. By (\ref{independence}), these depend on
at most two additional registers of qubits, which encode the two spine
labels to the left of those being acted upon. Thus, for any of the
standard Dehn twist generators, $U_{i,j}$ is a controlled unitary
acting on at most $5 \beta$ target qubits and $2\beta$ control
qubits. Recalling that $\beta = O(\log |g|)$, we can apply the
standard construction from section 4.5 of~\cite{Nielsen_Chuang} to
implement this unitary transformation with $\textrm{poly}(|g|)$
quantum gates, provided each matrix element of $U_{i,j}$ can be
computed efficiently. By (\ref{udef}), one sees that the only
potentially difficult part of computing the matrix elements of
$\ref{udef}$ is the computation of the thresholds. An efficient
classical algorithm for this task is given in appendix
\ref{thresholdcomp}.

\section{Hardness}

In this section we prove that the problem of estimating the normalized
WRT Fibonacci invariant of a mapping torus, given by a
polynomial-length word in the standard Dehn twist generators of the
mapping class group, to within $\pm \epsilon$ is DQC1-hard for
$\epsilon < 1/3900$. Generalizing our hardness proof beyond the
Fibonacci model seems less straightforward than generalizing our
algorithm. However, we consider it likely to be  
possible. Extending hardness to larger values of $\epsilon$ we leave
as an open problem. To prove hardness, we reduce from the problem of
estimating the absolute value of the normalized trace of a quantum
circuit. A proof of the hardness of absolute trace estimation is given
in appendix \ref{abs}. We thus require an efficient procedure that,
given a description of a quantum circuit for implementing a unitary
$U$, outputs a description of a mapping torus (i.e., a word in the
Dehn twist generators) whose WRT invariant is close to the trace of
$U$. It turns out to be convenient to suppose that $U$ is a quantum
circuit acting on a collection of 5-dimensional qudits
(``qupents''). As shown in appendix \ref{qudits}, this makes no
difference: the one-clean-qubit model is equivalent to the
one-clean-qupent model.

Let $U$ be a quantum circuit of $G$ gates acting on $n$
qupents arranged in a line. Without loss of generality, we may assume
that each gate acts either on a single qupent or a pair of neighboring
qupents. To prove hardness, we first define a many-to-one encoding
$\psi:S_{3n} \to \{0,1,2,3,4\}^n$, where $S_{3n}$ is the
set of fusion-consistent labelings of the standard spine of the
surface of genus $3n$. We divide the genus-$3n$ surface into $n$
segments, each having three handles. The number of fusion-consistent
labelings for a genus-three segment with two punctures depends on the
labels on the incoming and outgoing edges, as shown below.
\begin{center}
\includegraphics[width=0.36\textwidth]{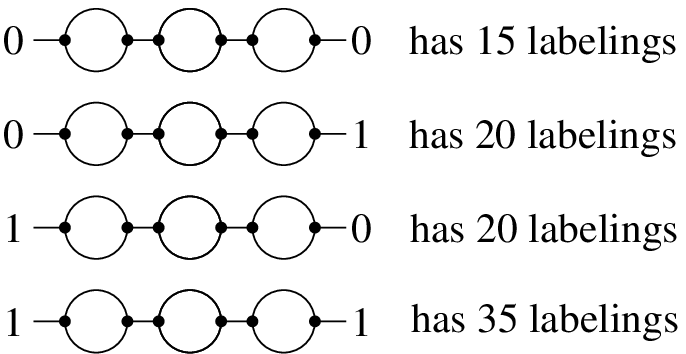}
\end{center}
In all cases, the number of fusion-consistent labelings is a multiple
of five. Thus, in every case a qupent can be encoded in the space of
labelings, together with a ``gauge'' qudit, whose value we ignore,
which has dimension 3, 4, or 7, depending on the labels of the incoming
edges. Thus the size $|\psi^{-1}(z)|$ of the preimage of any $z \in
\{0,1,2,3,4\}^n$ is exactly independent of $z$. Given any unitary $U$
acting on $n$ qupents, there corresponds a unitary acting on the span
of $S_{3n}$ which acts as $U$ on the encoded qupents, and as the
identity on the gauge qudits. We call this the $\psi$-encoding of $U$.

As shown in \cite{FLW02}, the Fibonacci representation of
the mapping class group of the genus $g>1$ surface is dense in the
corresponding unitary group, modulo phase. Thus, given any unitary
operation on $n$ qupents, we can find a sequence of Dehn twists which
approximates its $\psi$-encoding arbitrarily closely. The trace of the
$\psi$-encoding is thus equal to the trace of the original quantum
circuit, up to a phase. The remaining question is whether this
reduction can be done efficiently.

Cutting the genus-$3n$ surface into $n$ equal segments yields $n-2$
genus-3 doubly-punctured surfaces, and two genus-3 singly-punctured
surfaces, as shown below.
\begin{center}
\includegraphics[width=0.65\textwidth]{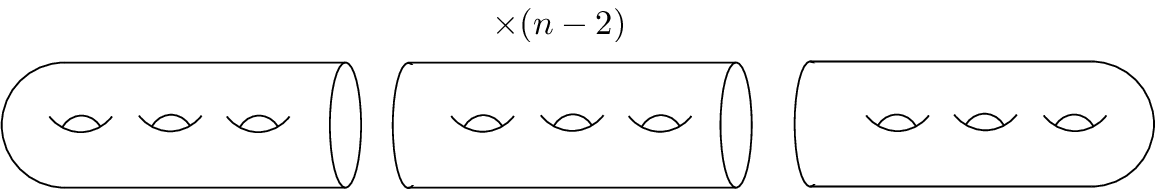}
\end{center}
One can pants-decompose a punctured surface, thereby associating the
surface to a spine. The spine has one ``external''edge for each
puncture, which attaches to the rest of the spine at only one
vertex. Upon labeling the spine, we can associate the label of any
external edge with the corresponding puncture. The Fibonacci
representation may then be extended to the label-preserving mapping
class group of the punctured surface. This group includes all the
standard Dehn twists, together with braiding of punctures with the
other punctures of the same label. In the Fibonacci representation,
braiding of zero-labeled punctures has no effect, thus a zero-labeled
puncture is equivalent to the absence of a puncture.

Theorem 6.2 of \cite{FLW02} states that for any fixed labels on the
punctures, the Fibonacci representation of the label-preserving
mapping class group of the $r$-punctured genus-$g$ surface is dense in
the corresponding unitary group modulo phase, provided $g+r >
1$. Thus, given any one-qupent gate, the Solovay-Kitaev
theorem~\cite{Nielsen_Chuang} efficiently yields a sequence of Dehn
twists and braid moves on the corresponding genus-3 singly-punctured
or doubly-punctured surface, whose Fibonacci representation
approximates the $\psi$-encoded gate arbitrarily closely. Similarly,
one efficiently approximates two-qupent gates by moves on genus-6
surfaces with one or two punctures.

We must modify the above construction so as not to use any braiding of
punctures. On the leftmost or rightmost qupents there is no problem;
the corresponding surfaces have only one puncture, and therefore
theorem 6.2 implies density without using any braiding
moves. Similarly, on any of the central surfaces, theorem 6.2 implies
density without using any braiding moves if at least one of the
punctures has a zero label. We can ensure this prior to the
application of any given gate by adapting the ``inchworm'' technique
from~\cite{Shor_Jordan}, as described in appendix \ref{inchworm}. In
this method, we bring a pair of zero labels adjacent to the target
segment, then implement the desired gate there, and carry the zeros to
the segment where the next gate is to be implemented. At the end, we
return these zeroes to their original location among the leftmost six
handles. As discussed in appendix \ref{inchworm}, the inchworm
construction entails some overhead in $\epsilon$, which gives rise to
the value $1/3900$.

In the above construction, we need density on two-punctured segments
in which one puncture is guaranteed to be labeled zero, and the other
puncture has unknown label. Theorem 6.2 of \cite{FLW02} implies
density separately in the subspace in which the other label is zero
and in which the other label is one. Because these subspaces have
different dimension (20 and 15, respectively) we may apply the
decoupling lemma from \cite{Aharonov_Arad}, which shows that a
sequence of Dehn twists can be found to approximate arbitrary pairs of
independent unitaries on these two subspaces, as desired.

\section{Analogy with Jones polynomials}
\label{sec:analogy}

In this paper we have shown that estimating the Turaev-Viro invariant
of a mapping torus in the Fibonacci model is
DQC1-complete. In~\cite{AJKR}, it was shown that estimating the
Turaev-Viro invariant of a general 3-manifold presented as a Heegaard
splitting is BQP-complete. Similarly, estimating the Jones polynomial
of the trace closure of a braid is DQC1-complete~\cite{Shor_Jordan,
  Jordan_Wocjan}, while estimating the Jones polynomial of the plat
closure of a braid is BQP-complete~\cite{AJL, Aharonov_Arad,
  Wocjan_Yard, Freedman1, Freedman2}. This suggests a relationship
between trace closures and mapping tori on one hand, and between plat
closures and Heegaard splittings on the other. Indeed, such a
relationship can be understood in the framework of axiomatic
topological quantum field theory, and suggests further generalizations
to, for instance, topological invariants of higher dimensional
manifolds.

\begin{figure}
\begin{center}
\includegraphics[width=0.5\textwidth]{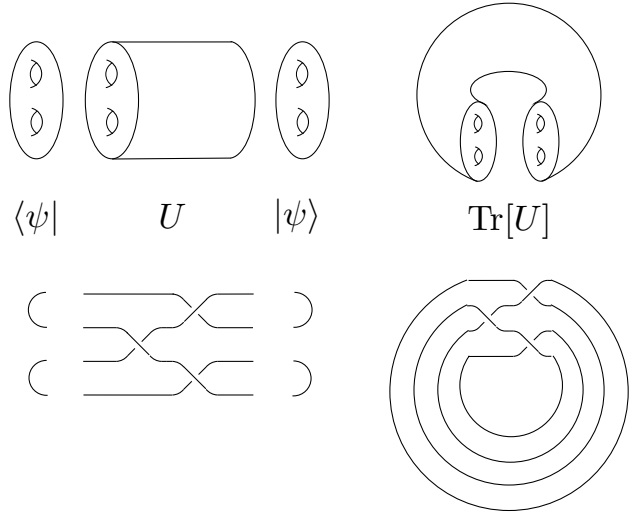}
\caption{\label{analogy}The problems of estimating the Jones
  polynomial of the plat closure of a braid and the Turaev-Viro
  invariant of a Heegaard splitting (left) are BQP-complete. The
  problems of estimating the Jones polynomial of the trace closure of
  a braid and the Turaev-Viro invariant of a mapping torus (right) are
  DQC1-complete. These situations are fundamentally analogous, as
  discussed in section \ref{sec:analogy}. We stress that the manifold
  figures are illustrations of the topological ideas behind this
  analogy, and are not correct two-dimensional projections of the
  manifolds themselves. In particular, after gluing, the two manifolds
  shown do not in reality have any boundaries.}

\end{center}
\end{figure}

\begin{figure}[ht]
\begin{center}
\includegraphics[width=0.4\textwidth]{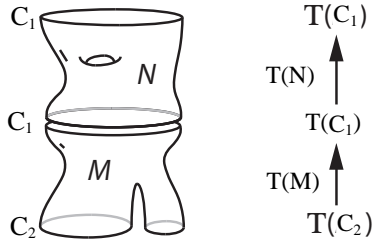}
\caption{\label{cobordisms}$M$ can be viewed as a two-manifold with two
  boundaries: a circle and a pair of circles. The TQFT associates a
  Hilbert space $T(C_2)$ to the pair of circles, a Hilbert space
  $T(C_1)$ to the circle, and a linear transformation $T(M):T(C_2) \to
  T(C_1)$ to $M$. Similarly, $T(N)$ is a linear transformation from
  $T(C_1)$ to itself. If we glue together $M$ and $N$ along the circle
  as shown, we obtain a manifold $MN$ with boundaries $C_2$ and
  $C_1$. The corresponding linear transformation is $T(NM) = T(N)
  \circ T(M)$.}
\end{center}
\end{figure}

A topological quantum field theory can be axiomatized as a functor $T$
from the category of cobordisms between $n$-manifolds to the category
of linear transformations between vector spaces~\cite{Atiyah,
  Walker}. That is, to each $n$-manifold the TQFT associates a vector
space, and to any $(n+1)$-manifold whose boundary is the union of two
disjoint $n$-manifolds the TQFT associates a linear transformation
between the two associated vector spaces. The functorial property means
that gluing together two cobordisms and then applying $T$ yields
the same linear transformation that is obtained by applying $T$ to
each of the two cobordisms and then composing the resulting linear
transformations; see figure \ref{cobordisms}. A TQFT maps the empty
$n$-manifold to the base field, which for the examples we consider is
always $\mathbb{C}$. Hence, for $M$ a manifold whose boundary
$\partial M$ has a single connected component, $T(M)$ is a map either
from $\mathbb{C}$ to the vector space $T(\partial M)$, that is, a
vector in $T(\partial M)$, or a map from $T(\partial M)$ to
$\mathbb{C}$, that is, a dual vector. The choice between these two
possibilities is determined by the orientation of the cobordism.

Recall that the genus-$g$ handlebody is the $3$-manifold whose
boundary is the genus-$g$ surface $\Sigma_g$. For example, the genus-1
handlebody is simply the solid donut. After assigning an orientation,
we may think of a handlebody as a cobordism from the empty manifold to
$\Sigma_g$, or as a cobordism from $\Sigma_g$ to the empty
manifold. Hence, in the TQFT framework, genus-$g$ handlebodies are
associated to vectors or dual vectors. We denote these as
$\ket{\psi_g}$ and $\bra{\psi_g}$, respectively. These vectors live in
the Hilbert space which the TQFT associates to $\Sigma_g$. In the case
of the Fibonacci model, this is precisely the vector space defined in
Section \ref{WRTdefinition}.

In the Fibonacci model, a cobordism from a surface to itself is mapped
to a unitary linear transformation $U$ on the associated Hilbert
space\footnote{We may think of the cobordism as describing a sort of
  spacetime evolution, while the unitary transformation describes the
  corresponding quantum time evolution. Indeed, this was one of the
  central motivating ideas behind the development of TQFTs.}. If the
surface is $\Sigma_g$, then we may ``cap'' the cobordism with
handlebodies on both ends. The resulting $3$-manifold has no boundary,
and thus corresponds to a linear map from $\CC$ to itself, i.e., a
complex number. In this case, this number is the matrix element
$\bra{\psi_g} U \ket{\psi_g}$, as illustrated in figure
\ref{analogy}. The problem of estimating a matrix element of the
unitary transformation induced by a quantum circuit is BQP-complete,
and this fact underlies the BQP-completeness proof for the Turaev-Viro
invariant of Heegaard splittings in~\cite{AJKR}. Instead of
``capping'' the two ends of the cobordism with handlebodies, we could
have simply glued the two ends together, resulting in a mapping
torus. This is again a $3$-manifold without boundary, which thus also
corresponds to a complex number. In a TQFT, gluing the two ends of a
cobordism corresponds to contracting the two indices of the linear
transformation. In other words, instead of a single matrix entry, we
now obtain the trace of $U$. Finding the trace of the unitary
transformation induced by a quantum circuit is DQC1-complete, and this
fact underlies the DQC1-completeness proof for the Turaev-Viro
invariant of mapping tori given in this paper.

The situation regarding Jones polynomials is directly
analogous. A TQFT gives us a unitary representation of the braid
group. Gluing the two ends of a braid together (i.e., taking
the trace closure), as illustrated on the righthand side of figure
\ref{analogy}, corresponds to taking the trace of the unitary and
yields a DQC1-complete problem. Caps correspond to vectors and dual
vectors depending on orientation, hence capping a braid (taking the
plat closure, as illustrated on the lefthand side of figure
\ref{analogy}) yields a matrix element of the associated unitary
transformation, and corresponds to a BQP-complete problem. The analogy
can be tightened further by noting that the braid group $B_n$ is
simply the mapping class group of the surface of genus zero and $n+1$
punctures (that is, the $n$-punctured disk), whereas in the case
of 3-manifold invariants we consider the mapping class group of the
genus-$g$ surface with no punctures. On the other hand, it is worth
bearing in mind that the notion of equivalence captured by the Jones
polynomial is ambient isotopy, in contrast to the Turaev-Viro and WRT
invariants, which capture homeomorphism.

The analogy presented here naturally suggests an extension of
BQP-completeness and DQC1-completeness results to $n$-manifold
invariants arising from TQFTs at higher $n$. More generally, one could
attempt to isolate a property of pairs, consisting of a group $G$ and
one of its representations $U$, such that estimating matrix entries of
$U$ is BQP-complete while estimating the trace of $U$ is
DQC1-complete. Perhaps one could find a general theorem encompassing
many such results. We leave this as an open problem.

\section{Acknowledgments}

We thank Robert K\"onig, Ben Reichardt and Edgar Bering for useful
discussions and some diagrams. S.J. acknowledges support from the
Sherman Fairchild Foundation and NSF grant
PHY-0803371. G.A. acknowledges support from NSERC, MITACS and ARO.

\appendix

\section{Equivalence Between One Clean Qudit Models}
\label{qudits}

Given a quantum circuit on $a$-dimensional qudits we wish to construct
a quantum circuit on $b$-dimensional qudits that has the same
trace. If $b=ca$ for some integer $c$ then this is easy. We just
consider each $b$-dimensional qudit to be an $a$-dimensional qudit
plus a $c$-dimensional ``gauge'' qudit that we ignore. Similarly, if
$b^d = ca$ for some integers $d,c$ then we can treat $d$-tuples of
$b$-dimensional qubits as an $a$-dimensional qudit plus a
$c$-dimensional gauge qudit. For these encodings, the encoded circuit
is easy to construct gate by gate. Given a gate acting on $n$
$a$-dimensional qudits, we can write down a unitary acting on $dn$
$b$-dimensional qudits equal to the original gate tensored with the
$c$-dimensional identity on the gauge system. This $dn$-dimensional
gate can be exactly decomposed into a product of $O(b^{2dn})$ 2-qudit
gates using the standard construction from section 4.5
of~\cite{Nielsen_Chuang}. Because $d$ and $n$ are constants, this is
sufficiently efficient. The normalized trace of the encoded circuit is
exactly equal to the normalized trace of the original circuit.

The harder case is when there do not exist integers $c$ and $d$ such
that $b^d = ca$. In this case we find $c,d \in \mathbb{Z}$ such that
$b^d \simeq ca$. Specifically, suppose we achieve
\begin{equation}
\label{upperlower}
\frac{ca}{b^d} = 1 - \delta
\end{equation}
for some $\delta \ll 1$. Then we can encode one $a$-dimensional qudit
plus a $c$-dimensional gauge qudit into $d$ $b$-dimensional qudits
with a few (namely $\delta b^d$) noncoding states left over. We can
define our encoded gates to act as the identity on these noncoding
states. If we make sure the noncoding states are a small fraction of
all $b^{dn}$ states, the normalized trace of the encoded circuit will
approximately match the normalized trace of the original circuit.

Let $U_a$ be the original unitary acting on $n$ $a$-dimensional qudits
and let $U_b$ be the unitary acting on $dn$ $b$-dimensional qudits, in
which we encode $U_a$ as described above. Then, $U_b$ acts on $b^{dn}$
states, of which $(ca)^n$ encode states of the original circuit,
\[
\frac{\tr [U_b]}{b^{dn}} = \frac{c^n \tr [U_a] +
  (b^{dn}-(ca)^n)}{b^{dn}}.
\]
The magnitude of the discrepancy $\Delta$ between the normalized
traces of $U_b$ and $U_a$ is thus
\begin{eqnarray*}
\Delta & = & \left| \frac{c^n \tr [U_a] +
  (b^{dn}-(ca)^n)}{b^{dn}} - \frac{\tr [U_a]}{a^n} \right| \\
& = & \left| \left( \left( \frac{ca}{b^d} \right)^n - 1 \right)
  \frac{\tr [U_a]}{a^n} + 1 - \left( \frac{ca}{b^d} \right)^n \right|
  \\
& \leq & \left| \left( \frac{ca}{b^d} \right)^n - 1 \right| \cdot
  \left| \frac{\tr [U_a]}{a^n} \right| + \left|1-\left( \frac{ca}{b^d}
  \right)^n \right| \\
& \leq & \left| \left( \frac{ca}{b^d} \right)^n - 1 \right| + \left| 1
  - \left( \frac{ca}{b^d} \right)^n \right| \\
& = & 2 \left| (1 - \delta)^n - 1 \right|.
\end{eqnarray*}
Thus if
\begin{equation}
\label{deltep}
\delta = \frac{\epsilon}{n}
\end{equation}
we have, for small $\epsilon$,
\begin{equation}
\label{limit}
\lim_{n \to \infty} \Delta = 2 \left| e^{-\epsilon} -1 \right| \simeq
2 \epsilon.
\end{equation}
Comparing (\ref{upperlower}), (\ref{deltep}), (\ref{limit}), we see
that in the limit of large $n$ and small $\epsilon$, in order to
achieve error upper bounded by $\Delta$ it suffices to obtain
\[
\frac{b^d - ca}{b^d} \leq \frac{\Delta}{2n}.
\]
For given $b,d,a$ there always exists an integer $c$ such that $b^d -
c \leq a$. So we just need to choose $d$ sufficiently large that
\[
\frac{a}{b^d} \leq \frac{\Delta}{2n}.
\]
Equivalently,
\[
d \geq \log_b \left( \frac{2na}{\Delta} \right).
\]

A $k$-qudit gate from $U_a$ thus gets encoded as a $dk$-qudit gate in
$U_b$. This encoded gate acts on a $b^{dk}$-dimensional space. We have
just shown that it suffices to choose $d = \log_b \left(
\frac{2na}{\Delta} \right)$. Thus the encoded $k$-qudit gate acts on a
$\left( \frac{2na}{\Delta} \right)^k$-dimensional space. Using the
construction from section 4.5 of~\cite{Nielsen_Chuang}, we can
implement an arbitrary $D$-dimensional unitary exactly with $O(D^2)$
2-qudit gates. Thus each $k$-qudit gate in $U_a$ gets encoded by $O
\left( \left( \frac{2na}{\Delta} \right)^{2k} \right)$ elementary
gates in $U_b$. By gate universality, we can assume $k \leq 2$, so our
encoding has an overhead quartic in $n$ and $1/\Delta$. This is
perhaps not very efficient, but is nevertheless polynomial, and
thus suffices to prove the equivalence of DQC1 defined with qudits of
any constant dimension.

\section{Estimating the Absolute Trace is DQC1-hard}
\label{abs}

In this section we slightly adapt the proof from \cite{Shepherd} to
show that estimating the absolute value of the trace of a quantum
circuit to within $\pm 1/24$ is a DQC1-complete problem. Consider an
arbitrary DQC1 computation. We start with the state $\ket{0} \bra{0}
\otimes \frac{\id}{2^n}$, apply an arbitrary quantum circuit $U$, and
then measure the first qubit in the $\ket{0},\ket{1}$ basis. Changing
the initial state of the pure qubit, or changing the measurement basis
does not add generality, as these changes can be subsumed into
$U$. The probability of measurement outcome $\ket{0}$ is
\begin{equation}
p_0 = \tr \left[ (\ket{0} \bra{0} \otimes \id) U (\ket{0} \bra{0}
  \otimes \id/2^n)U^\dag \right].
\end{equation}
Let $U'$ be the unitary implemented by the following quantum
circuit on $n+2$ qubits.
\[
U' = \begin{array}{l} \Qcircuit @C=1em @R=.7em {
 &  \qw   & \multigate{1}{U^\dag} & \ctrl{2} & \multigate{1}{U} & \ctrl{3} & \qw    & \qw \\
 & {/}\qw & \ghost{U^\dag}        & \qw      & \ghost{U}        & \qw      & {/}\qw & \qw \\
 &  \qw   & \qw                   & \targ    & \qw              & \qw      & \qw    & \qw \\
 &  \qw   & \qw                   & \qw      & \qw              & \targ    & \qw    & \qw
} \end{array}
\]
Thus, $p_0 = 2 \frac{\tr U'}{2^{n+2}}$, as one can see
by writing out the trace as a sum over diagonal matrix elements in the
computational basis. Because $p_0$ is real it is also true that  $p_0
= 2 \frac{\left| \tr U' \right|}{2^{n+2}}$. Hence
estimating the absolute value of the normalized trace of quantum
circuits to suffices to predict the outcome of any DQC1 experiment.

As is standard in the complexity theory of probabilistic computation,
``yes'' instances of DQC1 are defined to have acceptance probability
2/3 and ``no'' instances are defined to have acceptance probability
1/3. Thus, deciding DQC1 is equivalent to estimating the normalized
trace of a quantum circuit to within $\pm 1/6$. The reduction here has
a factor of four overhead in normalization, thus estimating the
absolute trace to within $\pm 1/24$ is DQC1-complete.

\section{Efficiently Computing Thresholds}
\label{thresholdcomp}

Consider the standard spine of the genus-$g$ surface, numbered as in
figure \ref{numbering}. Suppose edges 1 through $i$ have been
labeled in a fusion-consistent manner with anyon types
$s_1,\ldots,s_i$. We wish to compute how many completions there are to
this partial labelling. That is, we wish to compute the number of
fusion-consistent strings of $3g-3$ labels, whose first $i$ labels are
given by $s_1,\ldots,s_i$. 

Denote the horizontal edges of the standard spine from right to left
by $e_1,e_2,\ldots$, as shown below.
\begin{center}
\includegraphics[width=0.3\textwidth]{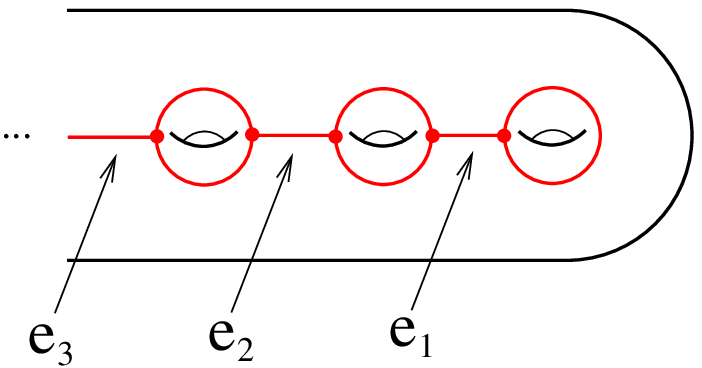}
\end{center}
Let $Z_b^{(k)}$ be the number of completions in which the rightmost
labeled edge is $e_k$ and has label $b \in \{0,1\}$. One sees that $Z_0^{(1)}
= 2$, and $Z_1^{(1)} = 1$, by the following enumeration of
fusion-consistent diagrams.
\begin{center}
\includegraphics[width=0.35\textwidth]{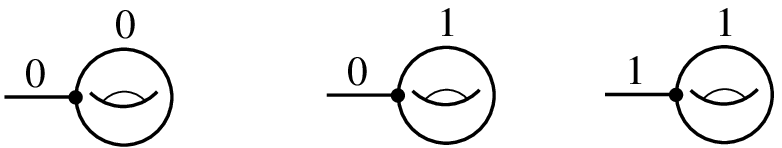}
\end{center}
Furthermore, we have the recurrence relations $Z_0^{(n+1)} = 2
Z_0^{(n)} + Z_1^{(n)}$ and $Z_1^{(n+1)} = 3 Z_1^{(n-1)} +
Z_0^{(n-1)}$, by the following enumeration of fusion-consistent
diagrams.
\begin{center}
\includegraphics[width=0.5\textwidth]{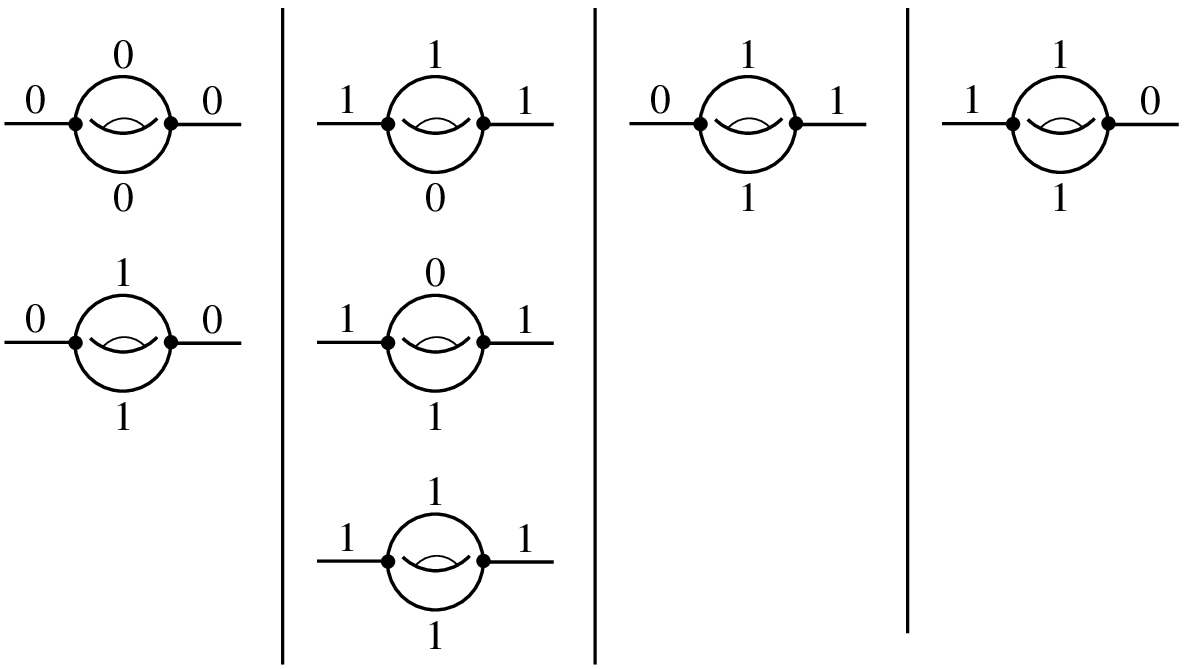}
\end{center}
Solving these recurrence relations yields
\[
\left[ \begin{array}{l} Z_0^{(n)} \\ Z_1^{(n)} \end{array} \right] =
\left[ \begin{array}{ll} 2 & 1 \\ 1 & 3 \end{array} \right]^{n-1}
\left[ \begin{array}{l} 2 \\ 1 \end{array} \right].
\]
The other two cases---completions starting with an upper curved
edge, or a lower curved edge---can be solved similarly. The $n\th$
power of a matrix may be computed using $O(\log n)$ operations, thus
calculating the number of completions for any $i$ in $O(\log g)$
steps. The corresponding thresholds are then immediately obtained by
taking ratios of these.

\section{Inchworm}
\label{inchworm}

Suppose the spine-labeling contains a segment of the following
form.
\begin{equation}
\label{bare_inchworm}
\includegraphics[width=0.3\textwidth]{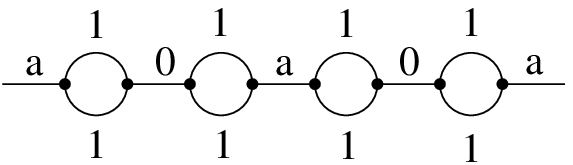}
\end{equation}
Here $a$ can be 1 or 0. We call this configuration the inchworm. We
may regard the right instance of $\begin{array}{c}
  \includegraphics[width=0.6in]{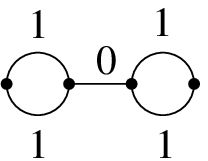} \end{array}$ as its head,
and the left instance as its tail. We next show a sequence of two
reversible operations by which we can move the inchworm one handle
rightward. In the first step the head moves one handle to the right,
leaving the tail in place, and in the second step, the tail catches
up, hence the name ``inchworm.''
\begin{center}
\includegraphics[width=0.36\textwidth]{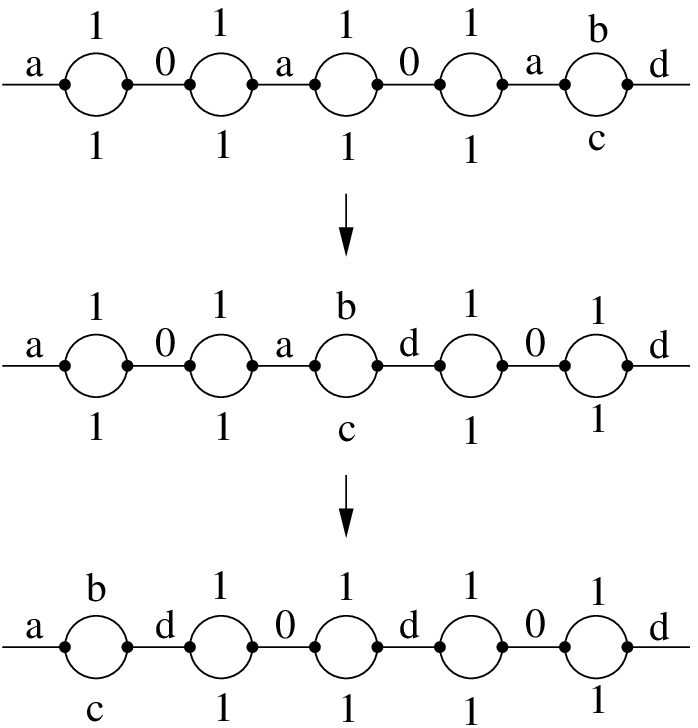}
\end{center}
Examination of the above diagram shows that if the fusion rules are obeyed
in the initial configuration, they are also obeyed in the intermediate
and final configurations. Furthermore, both steps are reversible
(\emph{i.e.} information preserving). Thus, they may be written as
permutation matrices acting on the space of allowed configurations,
and are therefore unitary. The first unitary transformation can be
implemented by local Dehn twists, because the zero in the tail of the
inchworm implies density of the Fibonacci representation on the
segment to the right of it. The second unitary transformation can be
implemented by local Dehn twists because the zero in the head of the
inchworm implies density on the segment to the left of it. (In both
steps, we are applying density to the twice-punctured genus-4 surface
with one puncture labeled zero. There are 75 labelings in which the
other puncture is labeled one and 50 labelings in which the other
puncture is labeled zero. Thus, the decoupling lemma of
\cite{Aharonov_Arad} implies density \emph{jointly} on these two
subspaces.) Repeating this process and its reverse, we may bring the
inchworm to any location within the spine.

To use the inchworm construction, we need to ensure that a segment of
the form (\ref{bare_inchworm}) exists in the first place. We may do
this by implementing a reversible operation on the leftmost six
handles, so that if the configuration (\ref{bare_inchworm}) is absent,
the matrix is strictly off-diagonal, and does not contribute to the
trace. Specifically, we consider the leftmost two handles to be an
ancilla system, and the next four handles to be the starting location
of the inchworm. If these four handles do not take the form
(\ref{bare_inchworm}) we cyclically permute the (five) basis states of the
ancilla system. Because this is done on the leftmost six handles, the
segment is only singly-punctured, and thus theorem 6.2 of \cite{FLW02}
implies density without braiding.

The noncontributing labelings decrease the normalized WRT by a
constant factor, which correspondingly necessitates decreasing the
precision parameter $\epsilon$ by the same factor. More precisely, in
the Fibonacci model, there are 325 fusion-consistent labelings for the
spine of the genus-four doubly-punctured surface. Among these, there
are two inchworm configurations ($a=0$ and $a=1$). Compounding this
$2/325$ normalization cost with the precision $\epsilon = 1/24$
obtained in appendix \ref{abs} for DQC1-hardness of absolute trace,
we find that estimating the normalized WRT invariant to within $\pm
1/3900$ is DQC1-hard.

As an aside, we note that the inchworm construction here is simpler
than that in \cite{Shor_Jordan}, in the following sense. The inchworm
construction of \cite{Shor_Jordan} involved reversible operations on
logarithmically large regions. Although the density theorems imply
that arbitrary reversible operations can be implemented on these
regions, they do not imply that the decomposition into local moves is
efficient. Rather this had to be explicitly proven in appendix D of
\cite{Shor_Jordan}. In contrast the inchworm construction here
involves reversible operations only on $O(1)$ handles, thus no
question of efficiency arises. 

\bibliography{maptorus}

\begin{thebibliography}{10}

\bibitem{Aharonov_Arad}
Dorit Aharonov and Itai Arad.
\newblock The {BQP}-hardness of approximating the {J}ones polynomial.
\newblock {\em New Journal of Physics}, 13:035019, 2011.
\newblock arXiv:quant-ph/0605181.

\bibitem{AJL}
Dorit Aharonov, Vaughan Jones, and Zeph Landau.
\newblock A polynomial quantum algorithm for approximating the {J}ones
  polynomial.
\newblock {\em STOC 06}, 2006.
\newblock arXiv:quant-ph/0511096.

\bibitem{AJKR}
Gorjan Alagic, Stephen Jordan, Robert K{\"o}nig, and Ben Reichardt.
\newblock Approximating {T}uraev-{V}iro 3-manifold invariants is universal for
  quantum computation.
\newblock {\em Physical Review A}, 82:040302(R), 2010.
\newblock arXiv:1003.0923.

\bibitem{Ambainis_Schulman_Vazirani}
Andris Ambainis, Leonard~J. Schulman, and Umesh Vazirani.
\newblock Computing with highly mixed states.
\newblock {\em Journal of the Association of Computing Machinery},
  53(3):507--531, 2006.
\newblock A preliminary version appears in 2000 and is available at
  arXiv:quant-ph/0003136.

\bibitem{Atiyah}
Michael Atiyah.
\newblock Topological quantum field theories.
\newblock {\em {Publications Math\'ematiques de l'IH\'ES}}, 68:175--186, 1988.

\bibitem{Brandao}
Fernando Brand{\~a}o.
\newblock {\em Entanglement Theory and the Quantum Simulation of Many-Body
  Physics}.
\newblock PhD thesis, Imperial College London, 2008.
\newblock arXiv:0810.0026.

\bibitem{Datta_thesis}
Animesh Datta.
\newblock {\em Studies on the Role of Entanglement in Mixed-state Quantum
  Computation}.
\newblock PhD thesis, University of New Mexico, 2008.
\newblock arXiv:0807.4490.

\bibitem{Datta_Flammia_Caves}
Animesh Datta, Steven~T. Flammia, and Carlton~M. Caves.
\newblock Entanglement and the power of one qubit.
\newblock {\em Physical Review A}, 72:042316, 2005.
\newblock arXiv:quant-ph/0505213.

\bibitem{Datta_Gharibian}
Animesh Datta and Sevag Gharibian.
\newblock Signatures of non-classicality in mixed-state quantum computation.
\newblock {\em Physical Review A}, 79:042325, 2009.
\newblock arXiv:0811.4003.

\bibitem{Datta_Shaji_Caves}
Animesh Datta, Anil Shaji, and Carlton~M. Caves.
\newblock Quantum discord and the power of one qubit.
\newblock {\em Physical Review Letters}, 100:050502, 2008.
\newblock arXiv:0709.0548.

\bibitem{Datta_Vidal}
Animesh Datta and Guifre Vidal.
\newblock On the role of entanglement and correlations in mixed-state quantum
  computation.
\newblock {\em Physical Review A}, 75:042310, 2007.
\newblock arXiv:quant-ph/0611157.

\bibitem{Freedman2}
Michael Freedman, Alexei Kitaev, and Zhenghan Wang.
\newblock Simulation of topological field theories by quantum computers.
\newblock {\em Communications in Mathematical Physics}, 227:587--603, 2002.
\newblock arXiv:quant-ph/0001071.

\bibitem{Freedman1}
Michael Freedman, Michael Larsen, and Zhenghan Wang.
\newblock A modular functor which is universal for quantum computation.
\newblock {\em Communications in Mathematical Physics}, 227:605, 2002.
\newblock arXiv:quant-ph/0001108.

\bibitem{FLW02}
Michael~H. Freedman, Michael~J. Larsen, and Zhenghan Wang.
\newblock The two-eigenvalue problem and density of {J}ones representation of
  braid groups.
\newblock {\em Communications in Mathematical Physics}, 228:177--199, 2002.

\bibitem{Garnerone}
S.~Garnerone, A.~Marzuoli, and M.~Rasetti.
\newblock Efficient quantum processing of three-manifold topological
  invariants.
\newblock {\em Advances in Theoretical and Mathematical Physics},
  13(6):1601--1652, 2009.
\newblock arXiv:quant-ph/0703037.

\bibitem{Jordan_Wocjan}
Stephen~P. Jordan and Pawel Wocjan.
\newblock Estimating {J}ones and {HOMFLY} polynomials with one clean qubit.
\newblock {\em Quantum Information and Computation}, 9, 2009.
\newblock arXiv:0807.4688.

\bibitem{Knill_Laflamme98}
E.~Knill and R.~Laflamme.
\newblock Power of one bit of quantum information.
\newblock {\em Physical Review Letters}, 81(25):5672--5675, 1998.
\newblock arXiv:quant-ph/9802037.

\bibitem{Knill_Laflamme01}
E.~Knill and R.~Laflamme.
\newblock Quantum computation and quadratically signed weight enumerators.
\newblock {\em Information Processing Letters}, 79(4):173--179, 2001.
\newblock arXiv:quant-ph/9909094.

\bibitem{Luo}
Shunlong Luo.
\newblock Using measurement-induced disturbance to correlations as classical or
  quantum.
\newblock {\em Physical Review A}, 77:022301, 2008.

\bibitem{Marx}
Raimund Marx, Amr Fahmy, Louis Kauffman, Samuel Lomonaco, Andreas Sp{\"o}rl,
  Nikolas Pomplun, John Myers, and Steffen~J. Glaser.
\newblock {NMR} quantum calculations of the {J}ones polynomial.
\newblock {\em arxiv:0909.1080}, 2009.

\bibitem{Nielsen_Chuang}
Michael~A. Nielsen and Isaac~L. Chuang.
\newblock {\em Quantum Computation and Quantum Information}.
\newblock Cambridge University Press, 2000.

\bibitem{Passante}
G.~Passante, O.~Moussa, C.~A. Ryan, and R.~Laflamme.
\newblock Experimental approximation of the {J}ones polynomial with {DQC1}.
\newblock {\em Physical Review Letters}, 103:250501, 2009.
\newblock arXiv:0909.1550.

\bibitem{Roberts}
Justin~D. Roberts.
\newblock Skein theory and {T}uraev-{V}iro invariants.
\newblock {\em Topology}, 34:771--787, 1995.

\bibitem{Shepherd}
Dan Shepherd.
\newblock Computation with unitaries and one pure qubit.
\newblock 2006.
\newblock arXiv:quant-ph/0608132.

\bibitem{Shor_Jordan}
Peter~W. Shor and Stephen~P. Jordan.
\newblock Estimating {J}ones polynomials is complete for one clean qubit.
\newblock {\em Quantum Information and Computation}, 8(8/9):681--714, 2008.
\newblock arXiv:0707.2831.

\bibitem{Turaev91}
V.~G. Turaev.
\newblock Topology of shadows.
\newblock preprint, 1991.

\bibitem{TuraevBook}
V.~G. Turaev.
\newblock {\em Quantum invariants of knots and 3-manifolds}, volume~18 of {\em
  de Gruyter studies in mathematics}.
\newblock de Gruyter, New York, 1994.

\bibitem{Walker}
Kevin Walker.
\newblock On {W}itten's 3-manifold invariants.
\newblock {\texttt{http://canyon23.net/math/1991TQFTNotes.pdf}}, 1991.

\bibitem{Wocjan_Yard}
Pawel Wocjan and Jon Yard.
\newblock The {J}ones polynomial: quantum algorithms and applications in
  quantum complexity theory.
\newblock {\em Quantum Information and Computation}, 8:147--180, 2008.
\newblock arXiv:quant-ph/0603069.

\end{thebibliography}

\end{document}